\documentclass[nofootinbib,showpacs,pra,preprint,aps,floatfix]{revtex4}
\usepackage{setspace}
\usepackage{bm}
\usepackage{amssymb}
\usepackage{graphicx}
\usepackage{mathtext}
\usepackage{amsfonts}
\usepackage{mathrsfs}
\usepackage{color}

\newcommand{\e}{\mathrm{e}}                  
\renewcommand{\d}[1]{\mathrm{d}#1}           
\renewcommand{\v}[1]{\mathbf{#1}}            

\newcommand{\wh}{\widehat}

\begin{document}

\title{Bohmian approach to spin-dependent time of arrival for particles in a uniform field and for particles passing through a barrier}

\author{S. V. Mousavi}
\email{s_v_moosavi@mehr.sharif.edu}
\affiliation{Department of Physics, Sharif University of Technology, P. O. Box 11365-9161, Tehran, Iran}
\affiliation{Institute for Studies in Theoretical Physics and Mathematics (IPM), P. O. Box 19395-5531, Tehran, Iran}
\author{M. Golshani}
\email{golshani@sharif.edu}
\affiliation{Department of Physics, Sharif University of Technology, P. O. Box 11365-9161, Tehran, Iran}
\affiliation{Institute for Studies in Theoretical Physics and Mathematics (IPM), P. O. Box 19395-5531, Tehran, Iran}

\begin{abstract}
It is known that Lorentz covariance fixes uniquely the current and the associated guidance law in the trajectory interpretation of quantum mechanics for spin-$\frac{1}{2}$ particles. In the nonrelativistic domain this implies a guidance law for electrons which differs by an additional spin-dependent term from the one originally proposed by de Broglie and Bohm. Although the additional term in the guidance equation may not be detectable in the quantum measurements derived solely from the probability density $\rho$, it plays a role in the case of arrival-time measurements. In this paper we compute the arrival time distribution and the mean arrival time at a given location, with and without the spin contribution, for two problems: 1) a symmetrical Gaussian packet in a uniform field and 2) a symmetrical Gaussian packet passing through a 1D barrier. Using the Runge-Kutta method for integration of the guidance law, Bohmian paths of these problems are also computed.
\end{abstract}

\maketitle

\section{introduction}
In classical mechanics, each particle follows a definite trajectory, and so it is clear what is meant by the time at which a particle arrives at a given place. In quantum mechanics, as opposed to classical physics, the meaning of the arrival time of a particle at a given location is not evident when the finite extent of the wave function and its spreading becomes relevant.  Moreover, in the quantum case one expects an arrival time distribution, and there are different proposals for this (see e.g. the review article in \cite{MuLe-PhysRep-2000}, and the book \cite{MuSaEg-book-2002}.) As pointed out by Hannstein et al \cite{HaHeMu-JPB-2005}, these arrival time distributions have been controversial, since they are derived from purely theoretical arguments without specifying a measurement procedure.
The lack of a self-adjoint arrival-time operator conjugate to the free Hamiltonian lies at the core of the difficulties in the formulation of quantum arrival times \cite{HeSeMu-PRA-2003}. In Bohm's interpretation of nonrelativistic quantum mechanics \cite{BoI-PR-1952, BoII-PR-1952, HiBo-book-1993, Ho-book-1993, DuGoZa-1992-JSP, DuGoZa-1996-book, Tu-2004-AJP}, an electron is a particle that has a well defined trajectory and so the definition of arrival-time distributions is unambiguous. In Bohm's trajectory approach to the calculation of various characteristic times, it is only the particle component of the particle that is being clocked. If the arrival-time detector is not included in the Hamiltonian, then one has an expression for the ideal or intrinsic arrival-time distribution \cite{MuSaEg-Le-book-2002}. The arrival-time problem is unambiguously solved in the Bohmian mechanics, where for an arbitrary scattering potential $V(\v{x})$, one finds \cite{Le-PhyslettA-1993, MuSaEg-Le-book-2002, Le-book-1996, Le-PRA-1998} for those particles that actually reach $\v{x}=\v{X}$; the arrival-time distribution being given by the modulus of the probability current density, i.e., $|\v{J}(\v{X}, T)|$. In nonrelativistic quantum mechanics, the form of the current density $\v{J}$ is not uniquely determined by the continuity equation, a point which has been mentioned by a number of authors \cite{DeGh-Found.phys-1998, Ho-Found.phys-1998, Fi-PRA-1999}. It is determined only up to a divergenceless vector. For instance, one can construct a new current $\v{J}^{\prime}$ by adding the divergenceless current $\v{J}_s$ to the current $\v{J}$. The newly defined current $\v{J}^{\prime} =\v{J} + \v{J}_s$ also satisfies the continuity equation, with the same probability density. Although the additional contribution in the guidance equation may not be detectable in quantum measurements derived solely from the probability density $\rho$, it plays a role in the case of arrival-time measurements. In the case of spin-$1/2$ particles, Holland \cite{Ho-PRA-1999} showed that the particle current in the relativistic spin-$1/2$ Dirac theory is unique. Demanding that the non-relativistic spin-$1/2$ particle current be obtained from the nonrelativistic limit of the Dirac current, this nonrelativistic current is also unique. Struyve et al \cite{StBaNeWe-PLA-2004} have considered the uniqueness of paths for spin-$0$ and spin-$1$ particles. The spin-dependent Bohm trajectories have been investigated for hydrogen eigenstates \cite{CoVr-PLA-2002} and an electronic transition in hydrogen \cite{CoVr-JPA-2003}.

It was argued that for free spin eigenstates, spin contributions would in principle be experimentally distinguishable for arrival-time distributions of spinless and spin-$1/2$ particles \cite{AlMaDiSh-PRA-2003}. Holland et al \cite{HoPh-PRA-2003} have explored some of the implications of the modified guidance law in the case of the two-slit quantum interference. In this  case, the trajectories cross each other and they also cross the symmetry axis.

 The aim of the present paper is to explore some of the implications of the revised nonrelativistic guidance equation. Sec. \ref{Sec: 2} contains a very brief review of relevant parts of Bohm's interpretation of quantum mechanics. In Sec. \ref{Sec: 3}, by using the nonrelativistic limit of the Dirac current density, numerical computations of the effect of the spin-dependent term on arrival time at a given location, are presented for both a Gaussian wave in a uniform field and a Gaussian wave passing through a barrier. 


\section{Bohm's trajectory interpretation of quantum mechanics and the arrival time distributions} \label{Sec: 2}

In nonrelativistic Bohmian mechanics the world is described by point-like particles which follow trajectories determined by a law of motion. The evolution of the positions of these particles are guided by a wave function which itself evolves according to the Schr\"odinger equation. Bohmian mechanics makes the same predictions as the ordinary nonrelativistic quantum mechanics for the results of any experiment, provided we assume a random distribution for configuration of the system and the apparatus at the beginning of the experiment, given by $\rho(\v{x}, 0)= \Psi^{\dagger}(\v{x}, 0) \Psi(\v{x}, 0)$. If the probability density for the  configuration satisfies $\rho(\v{x}, t_0)= \Psi^{\dagger}(\v{x}, t_0) \Psi(\v{x}, t_0)$ at some initial time $t_0$, then the density to which this is carried by the continuity equation at any time $t$ is also given by $\rho(\v{x}, t)= \Psi^{\dagger}(\v{x}, t) \Psi(\v{x}, t)$ ~\cite{Bo-PR-1953}. 
As most of the quantum measurements boil down to position measurements, Bohm's theory and the standard quantum mechanics generally yield the same detection probabilities. The situation is different, however, if one considers, for example, measurements involving time-related quantities, such as arrival times, tunnelling times, etc. Bohm's theory makes unambiguous predictions for such measurements, but there is no consensus about what these quantities should be in the conventional quantum mechanics~\cite{St-quant-2005}. Given the initial position $\v{x}^{(0)} \equiv \v{x}(t=0)$ of a particle with the initial wave function $\Psi(\v{x}, t=0)$, its subsequent trajectory $\v{x}(\v{x}^{(0)}, t)$ is uniquely determined by the simultaneous integration of the time dependent Schr\"odinger equation, and the guidance equation $\frac{d\v{x}(t)}{dt}=\v{v}(\v{x}(t), t)$, in which $\v{v}=\frac{\v{J}}{\rho}$. For spinless particles, the Schr\"odinger equation yields:

\begin{eqnarray} \label{eq: spinless current}
\v{J}(\v{x}, t) &=& (\hbar/m) Im \bigl( \psi^{\ast}(\v{x}, t) \v{\nabla} \psi(\v{x}, t) \bigr)~.
\end{eqnarray}
But, for systems with more than one spatial dimension the probability current density, and hence the particle equation of motion, is not uniquely defined within nonrelativistic quantum mechanics. But, Holland \cite{Ho-PRA-1999} has shown that the probability current density deduced from the continuity equation is uniquely defined for Dirac electrons, when Lorentz covariance is imposed. Taking the nonrelativistic limit for a spin eigenstate in the absence of a magnetic field, one gets

\begin{widetext}
\begin{eqnarray} \label{eq: current}
\v{J}(\v{x}, t; \wh{\v{s}}) &=& \v{J}(\v{x}, t) + (\hbar/2m) \bigl( \v{\nabla} \rho(\v{x}, t)\bigr) \times \wh{\v{s}}\nonumber\\
 &=& (\hbar/m)\bigl[ Im \bigl( \psi^{\ast}(\v{x}, t) \v{\nabla} \psi(\v{x}, t) \bigr) + Re \bigl( \psi^{\ast}(\v{x}, t) \v{\nabla} \psi(\v{x}, t) \bigr) \times  \wh{\v{s}} \bigr]~,
\end{eqnarray}
\end{widetext}
where

\begin{eqnarray}
\v{s} &=& \frac{\hbar}{2} \wh{\v{s}} = \frac{\hbar}{2} \chi^{\dagger} \wh{\v{\sigma}}\chi~,
\end{eqnarray}
is the spin vector associated with the spin eigenstate $\chi$, which is a two-component spinor normalized to unity ($\chi^{\dagger} \chi = 1$). The nonrelativistic particle is, in effect, guided by the two-component wave function $\Psi(\v{x}, t; \v{s}) \equiv \psi(\v{x}, t) \chi$, and the original expression for the nonrelativistic velocity field should then be replaced by

\begin{eqnarray} \label{eq: guidance law}
\v {v}(\v{x}, t; \wh{\v{s}}) &\equiv& \frac{\v{J}(\v{x}, t; \wh{\v{s}})}{|\psi(\v{x}, t)|^2}~.
\end{eqnarray}

 A striking property of the spin-dependent term is that the components of the particle motion in orthogonal directions are generally mutually dependent, even when the wave function factorizes in these direction. For a 3D system with a wave function which is in the factorized form $\psi(\v{x}, 0)= \psi_x(x, 0) \psi_y(y, 0) \psi_z(z, 0)$ at $t=0$, we have
\begin{eqnarray*}
\psi(\v{x}, t) &=& \psi_x(x, t) \psi_y(y, t) \psi_z(z, t),
\end{eqnarray*}
at $t>0$, and for $\wh{\v{s}}=(0, 0, 1)$, Eq. (\ref{eq: guidance law}) becomes

\begin{eqnarray} \label{eq: velocity}
v_x &=& \frac{\hbar}{m}~\bigl[ Im \bigl( \frac{\partial_x \psi_x}{\psi_x} \bigr) + Re \bigl( \frac{\partial_y \psi_y}{\psi_y} \bigr) \bigr]~, \nonumber\\
v_y &=& \frac{\hbar}{m}~\bigl[ Im \bigl( \frac{\partial_y \psi_y}{\psi_y} \bigr) - Re \bigl( \frac{\partial_x \psi_x}{\psi_x} \bigr) \bigr]~, \nonumber\\
v_z &=& \frac{\hbar}{m}~\bigl[ Im \bigl( \frac{\partial_z \psi_z}{\psi_z} \bigr) \bigr]~,
\end{eqnarray}
where we have used the notation $\partial_x \psi_x = \frac{d\psi_x}{dx}$ and so on.
%
 For a nonrelativistic guiding wave of the form $\Psi(\v{x}, t; \wh{\v{s}}) = \psi(\v{x}, t) \chi$, with $\psi(\v{x}, t)$ being a solution of the Schr\"odinger equation and $\chi$ being a fixed spinor, Bohm's trajectory result for the distribution of particle arrival times $T$, at a given location $\v{x}=\v{X}$, is given by \cite{MuSaEg-Le-book-2002} 

\begin{eqnarray}
\Pi(T, \v{X}; \wh{\v{s}}) &=& |\v{J} (\v{X}, T; \wh{\v{s}})| / \displaystyle \int_{0}^\infty \d t |\v{J} (\v{X}, t; \wh{\v{s}})|~.
\end{eqnarray}

The mean arrival time of the particles reaching a detector located at $\v{X}$ is given by

\begin{eqnarray} \label{eq: mean arrival time}
\tau &=& \displaystyle \int_{0}^\infty \d t~t~\Pi(t, \v{X}; \wh{\v{s}}) = \frac{\displaystyle \int_{0}^\infty \d t~t~|\v{J} (\v{X}, t; \wh{\v{s}})|}{\displaystyle \int_{0}^\infty \d t~ |\v{J} (\v{X}, t; \wh{\v{s}})|}~,
\end{eqnarray}
in the presence of the spin-dependent contribution and 

\begin{eqnarray} \label{eq: mean arrival time}
\tau_i &=& \displaystyle \int_{0}^\infty \d t~t~\Pi_i(t, \v{X}) = \frac{\displaystyle \int_{0}^\infty \d t~t~|\v{J}(\v{X}, t)|}{\displaystyle \int_{0}^\infty \d t~ |\v{J} (\v{X}, t)|}~,
\end{eqnarray}
in the absence of the spin-dependent contribution. It must be mentioned that the definition of the mean arrival time used in Eq. (\ref{eq: mean arrival time}) is not a uniquely derivable result within standard quantum mechanics. We compute the arrival time distribution and the mean arrival time at a given location for two problems: 1) a symmetrical Gaussian packet in a uniform field and 2) a symmetrical Gaussian packet passing through a 1D barrier with and without the spin contribution. Using the Runge-Kutta method for the integration of the guidance law, Bohmian paths of these problems, have been also computed.


\subsection{Symmetric Gaussian packet in a uniform field}

 For the potential $V(\v{x})=\v{K}.\v{x}$, the wave function and the probability density at time $t$, calculated from \cite{Ho-book-1993}, are given by:

\begin{widetext}
\begin{eqnarray} \label{Eq: uf wave}
\psi(\v{x}, t) &=& (2 \pi s_t^2)^{-3/4} exp\{-(\v{x}-\v{u}t+\v{K}t^2/2m)^2/4s_t\sigma_0 \nonumber\\ &+&(im/\hbar)[(\v{u}-\v{K}t/m).(\v{x}-\frac{1}{2}\v{u}t)-\v{K}^2t^3/6m^2]\}~, \nonumber\\
\rho(\v{x}, t) &=& (2 \pi \sigma^2)^{-3/2} exp{-(\v{x}-\v{u}t+\v{K}t^2/2m)^2/2\sigma^2}~,
\end{eqnarray}
\end{widetext}
provided that this symmetrical Gaussian wave function is centred around the origin at $t=0$. In the Eq. (\ref{Eq: uf wave}), $s_t=\sigma_0(1+i\hbar t/2m\sigma_0^2)$, $\sigma=\sigma_0[1+(\frac{\hbar t}{2m\sigma_0^2})^2]^{1/2}$ and $\v{u}$ is the group velocity. The Eq. (\ref{Eq: uf wave}) is in the factorized form in three coordinate directions. We choose the uniform field and the group velocity to be in the $x$ direction i.e., $V(\v{x})=\v{K}.\v{x}=Kx$ and $\v{u}=(u, 0, 0)$ and the spin vector in the $z$ direction. The speed of the wave packet's centre in the $y$ and $z$ directions has been set equal to zero and therefore this function will spread but will not propagate in these directions. Using the Eq. (\ref{eq: velocity}), one gets for this factorized wave

\begin{eqnarray}
v_x &=& u-\frac{Kt}{m}+\frac{\hbar^2 t}{4 m^2 \sigma_0^4+\hbar^2 t^2}(x-ut+\frac{Kt^2}{2m})-\frac{\hbar}{2\sigma^2}y~, \nonumber\\
v_y &=& \frac{\hbar^2 t}{4 m^2 \sigma_0^4+\hbar^2 t^2}y+\frac{\hbar}{2\sigma^2}(x-ut+\frac{Kt^2}{2m})~, \nonumber\\
v_z &=& \frac{\hbar^2 t}{4 m^2 \sigma_0^4+\hbar^2 t^2}z~,
\end{eqnarray}
from which one can compute Bohmian trajectories. In the $z$ direction, one gets $z(t)=z_0 \bigl(1+\frac{\hbar^2 t^2}{4 m^2 \sigma_0^4} \bigr)^{1/2}$, in which $z_0$ is the $z$-component of the initial position of the particle.


\subsection{Symmetric Gaussian packet passing through a barrier in the x-direction} 

 suppose $V(\v{x})=V(x)=V_0$ for $0 \leq x \leq d$ and zero otherwise. The wave function at time $t$ is given by 

\begin{eqnarray}
\psi_x(x, t) &=& \int\d k \phi_{k}(x) \tilde{\psi}(k) \e^{-i E_k t/\hbar}~, \nonumber\\
\psi_y(y, t) &=& \frac{1}{(2\pi s_t^2)^{1/4}} e^{-y^2/4 s_t\sigma_0}~, \nonumber\\
\psi_z(z, t) &=& \frac{1}{(2\pi s_t^2)^{1/4}} e^{-z^2/4 s_t\sigma_0}~,
\end{eqnarray}
in which
\begin{eqnarray}
\phi_{k}(x) &\equiv& \phi_{k,T}(x) = \frac{1}{\sqrt{2\pi}} T(k) \e^{ikx}~~~~~~~~~~~x \geq d \nonumber\\
\tilde{\psi}(k) &=& \bigl( \frac{2 \sigma_0^2}{\pi} \bigr)^{1/4} \e^{-\sigma_0^2 (k-k_0)^2} ~,
\end{eqnarray}
where $\v{u}=(\hbar k_0/m, 0, 0)$ is the group velocity and 
\begin{eqnarray}
T_k &=& e^{-ikd} \frac{2kq}{2kq cos(qd)-i(q^2+k^2) sin(qd)}~~~~~\hbar^2 k^2/2m \neq V_0~, \nonumber\\
T_k &=& \frac{2}{2+ikd} e^{-ikd}~~~~~~~~~~~~~~~~~~~~~~~~~~\hbar^2 k^2/2m = V_0~,
\end{eqnarray}
is the transmission probability with $q=\sqrt{2m(\hbar^2 k^2/2m-V_0)}/\hbar$. Again, the speed of the wave packet's centre in $y$ and $z$ directions has been set equal to zero and therefore this function will spread but not propagate in these directions. For the motion in the $z$ direction, one gets

\begin{eqnarray}
z(t)=z_0 \bigl(1+\frac{\hbar^2 t^2}{4 m^2 \sigma_0^4} \bigr)^{1/2}~, 
\end{eqnarray}
in which $z_0$ is the $z$-component of the initial position of the particle.

\section{Numerical Results} \label{Sec: 3}
Numerical calculations are presented for a case of symmetric Gaussian packet in a uniform field and a case of symmetric Gaussian packet passing through a barrier in the x-direction. We compute the mean arrival time and the arrival time distribution with and without the spin-dependent term. Using the Runge-Kutta method for the integration of guidance equation, we compute some Bohmian paths for each case.

\subsection{Symmetric Gaussian packet in a uniform field}

 The initial wave packet is peaked at $(x_0= 0, y_0= 0, z_0= 0)$, with $\sigma_0 = 5$\AA{}. The detector position is chosen at $(x= 20, y= 20, z= 20)$\AA{}. 

\begin{figure}
\centering
\includegraphics[width=10cm,angle=0]{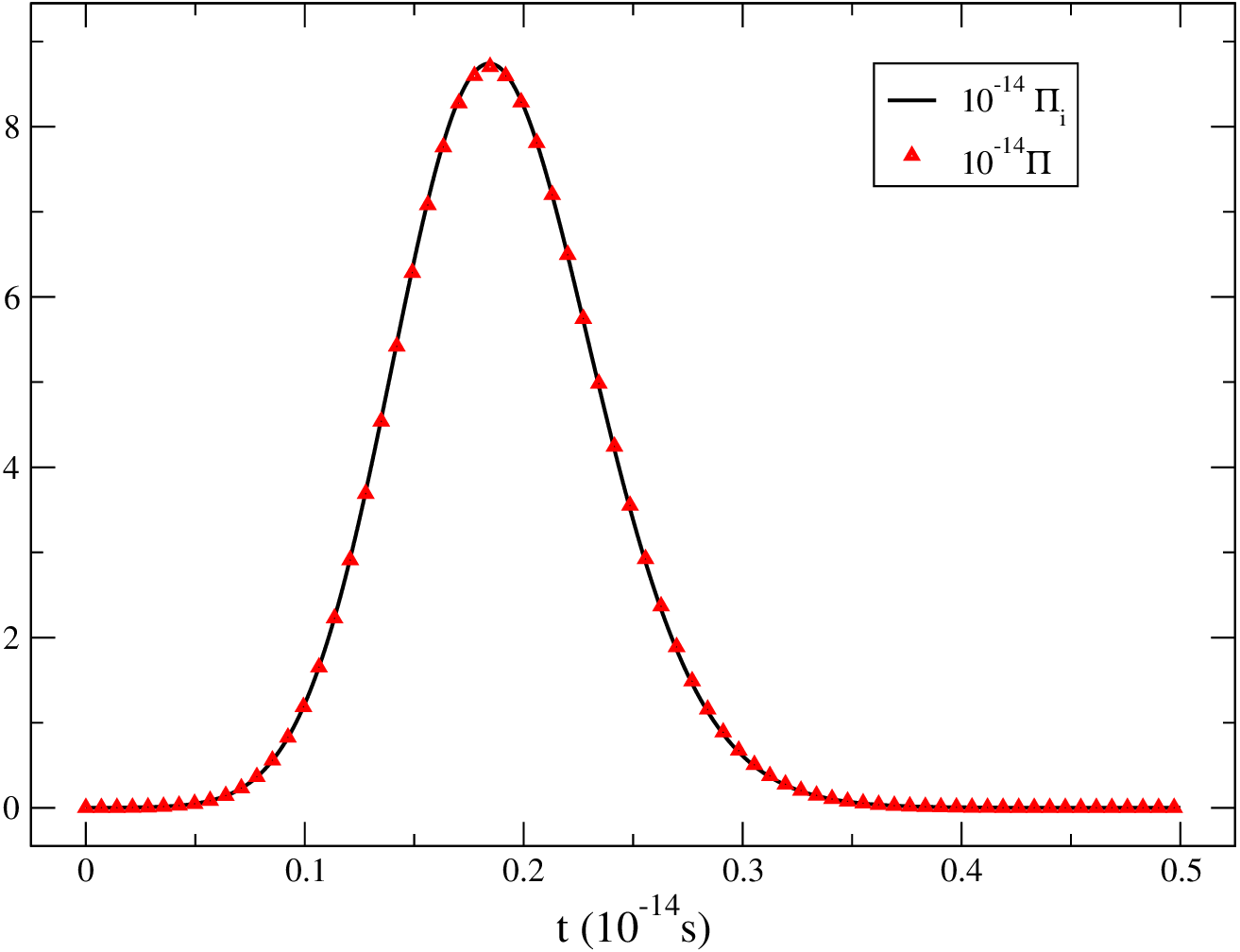}
\caption{The arrival time distribution of electrons versus time, without the spin-dependent term (solid curve) and with the spin-dependent term (shown by triangles), for a symmetric Gaussian packet in a uniform field.}
\vspace*{.6cm}
\label{fig: ar_dis_u}
\end{figure}

Fig. \ref{fig: ar_dis_u} shows the arrival time distribution of electrons for $K= mg$, corresponding to particles in a gravity field, with $E_0= \frac{1}{2} mu^2= 5$ev and $g= 9.8m/s^2$.
 
\begin{figure}
\centering
\includegraphics[width=10cm,angle=0]{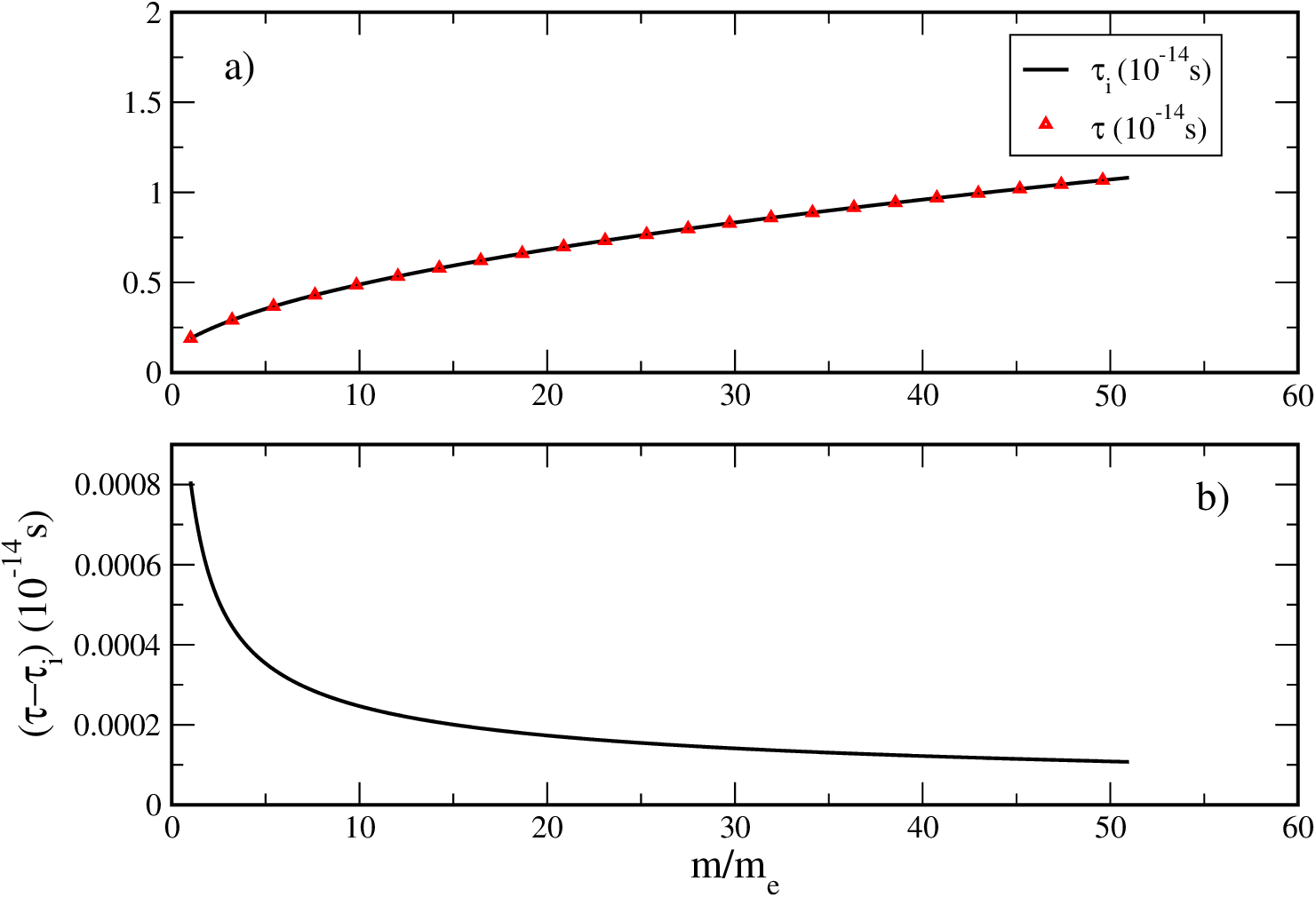}
\caption{a) The mean arrival time versus the ratio of the mass of the arriving particles to the mass of electron, without the spin-dependent term (solid curve) and with the spin-dependent term (shown by triangles) and b) the difference between $\tau_i$ and $\tau$ versus $m/m_e$ for a symmetric Gaussian packet in a uniform field.}
\label{fig: mean-mass-u}
\end{figure}

Fig. \ref{fig: mean-mass-u} shows the mean arrival time versus the mass of the arriving particle for a fixed value of $E_0$ ($E_0=5$ev). From this figure, it follows that the spin-dependent term increases the mean arrival time ($\tau > \tau_i$) and has very small effect on the mean arrival time at the given location, and the quantity $\tau-\tau_i$ is decreased with mass. 

\begin{figure}
\centering
\includegraphics[width=10cm,angle=0]{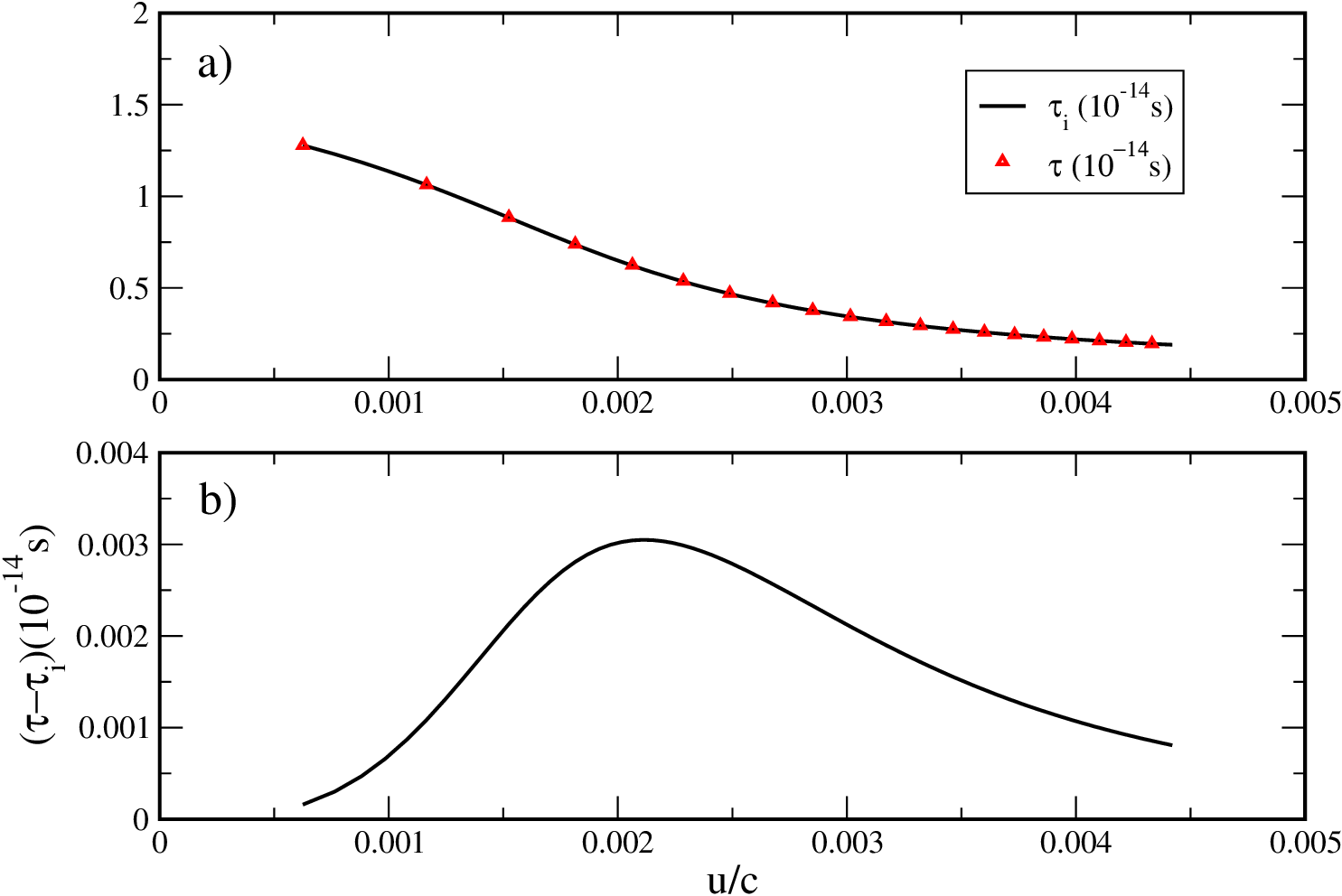}
\caption{a) The mean arrival time of electrons versus the ratio of the group velocity to the light velocity, without the spin-dependent term (solid curve) and with the spin-dependent term (shown by triangles) and b) the difference between $\tau_i$ and $\tau$ versus $u/c$ for a symmetric Gaussian packet in a uniform field.}
\label{fig: mean-speed-u}
\end{figure}

Fig. \ref{fig: mean-speed-u} shows the mean arrival time of electrons versus the ratio of the group velocity to the light velocity $u/c$ (Note that we have fixed the mass). It follows from this figure that the spin-dependent term increases the arrival time ($\tau > \tau_i$) and has very small effect on the mean arrival time at the given location and that the quantity $\tau-\tau_i$ is increased with the initial group velocity at first and then is decreased with it. 

\begin{figure}
\centering
\includegraphics[width=10cm,angle=0]{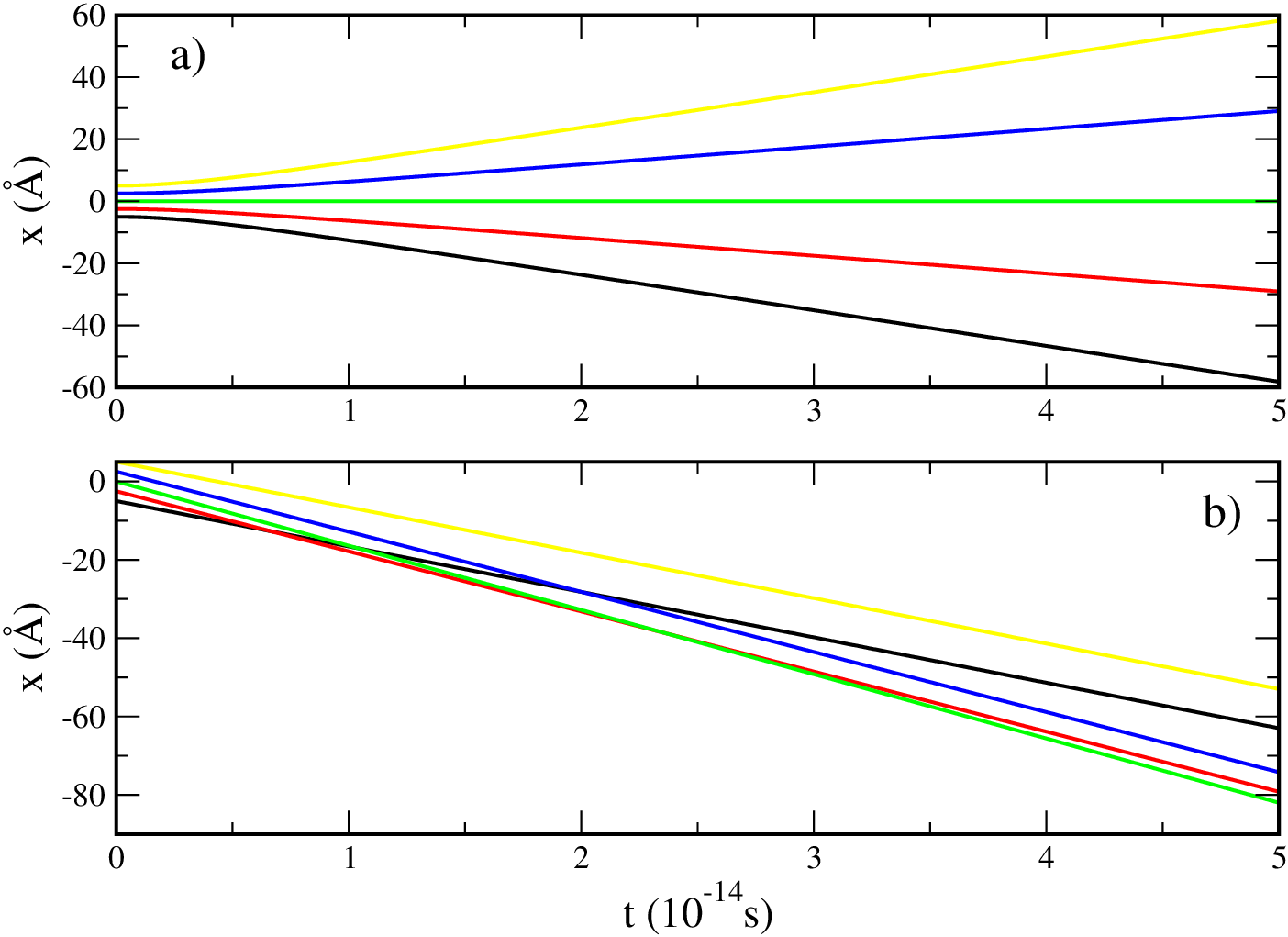}
\caption{Some Bohmian trajectories of electrons in x-t plane with the initial position $(x^{(0)}, y^{(0)}=\sqrt{2 \sigma_0^2-x^{(0)^2}},  z^{(0)}=0)$ a) in the absence of the spin-dependent term and b) in the presence of the spin-dependent term, for a symmetric Gaussian packet in a uniform field.}
\vspace*{.4cm}
\label{fig: tra_u}
\end{figure}

Fig. \ref{fig: tra_u} shows some Bohmian paths for electrons. One can see that some trajectories cross each other when one considers the spin-dependent term. It should be noted that trajectories cross each other for different values of the $y$ coordinate. Consequently there is no problem with the single-valuedness of the wave function. The spin-dependent term can also change the fate of the individual trajectories. There are some trajectories that don't reach the detector location, with the spin-dependent term, but reach the detector location without this term. 

\subsection{Symmetric Gaussian packet passing through a barrier in the x-direction}

 The initial wave packet is peaked at $(x_0= -10 \sigma_0, y_0= 0, z_0= 0)$ with $\sigma_0 = 5$\AA{}. The detector location  is chosen at $(x= 20, y= 20, z= 20)$\AA{}. 

\begin{figure}
\centering
\includegraphics[width=10cm,angle=0]{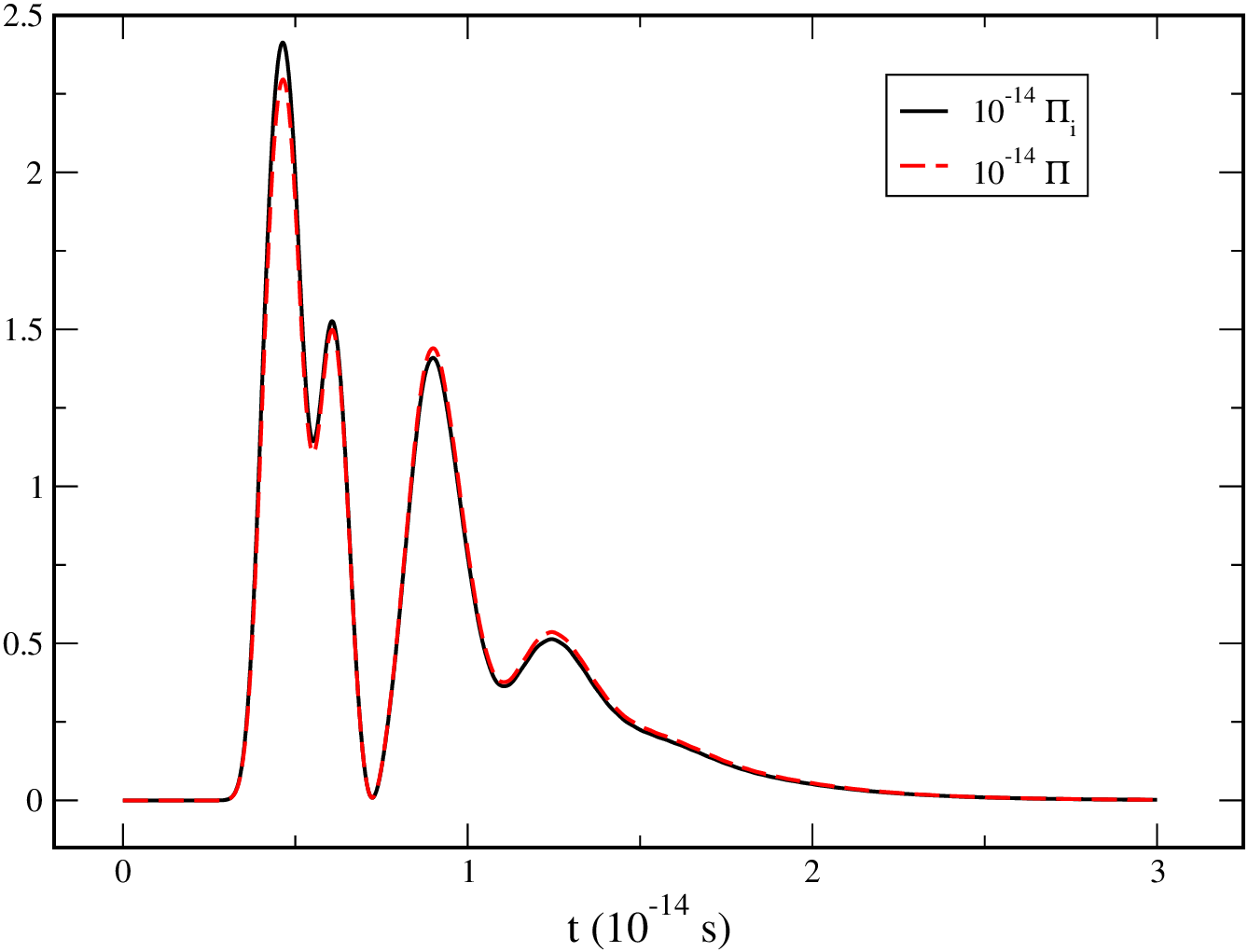}
\caption{The arrival time distribution of electrons versus time, without the spin-dependent term (solid curve) and with the spin-dependent term (dashed curve), for a symmetric Gaussian packet passing through a barrier in the x-direction.}
\label{fig: ar_dis_b}
\end{figure}

Fig. \ref{fig: ar_dis_b} shows the arrival time distribution of electrons, for $V_0= 8$ev, $E_0= \frac{1}{2} mu^2= 10$ev and $d= 10$\AA{}. 

\begin{figure}
\centering
\includegraphics[width=10cm,angle=0]{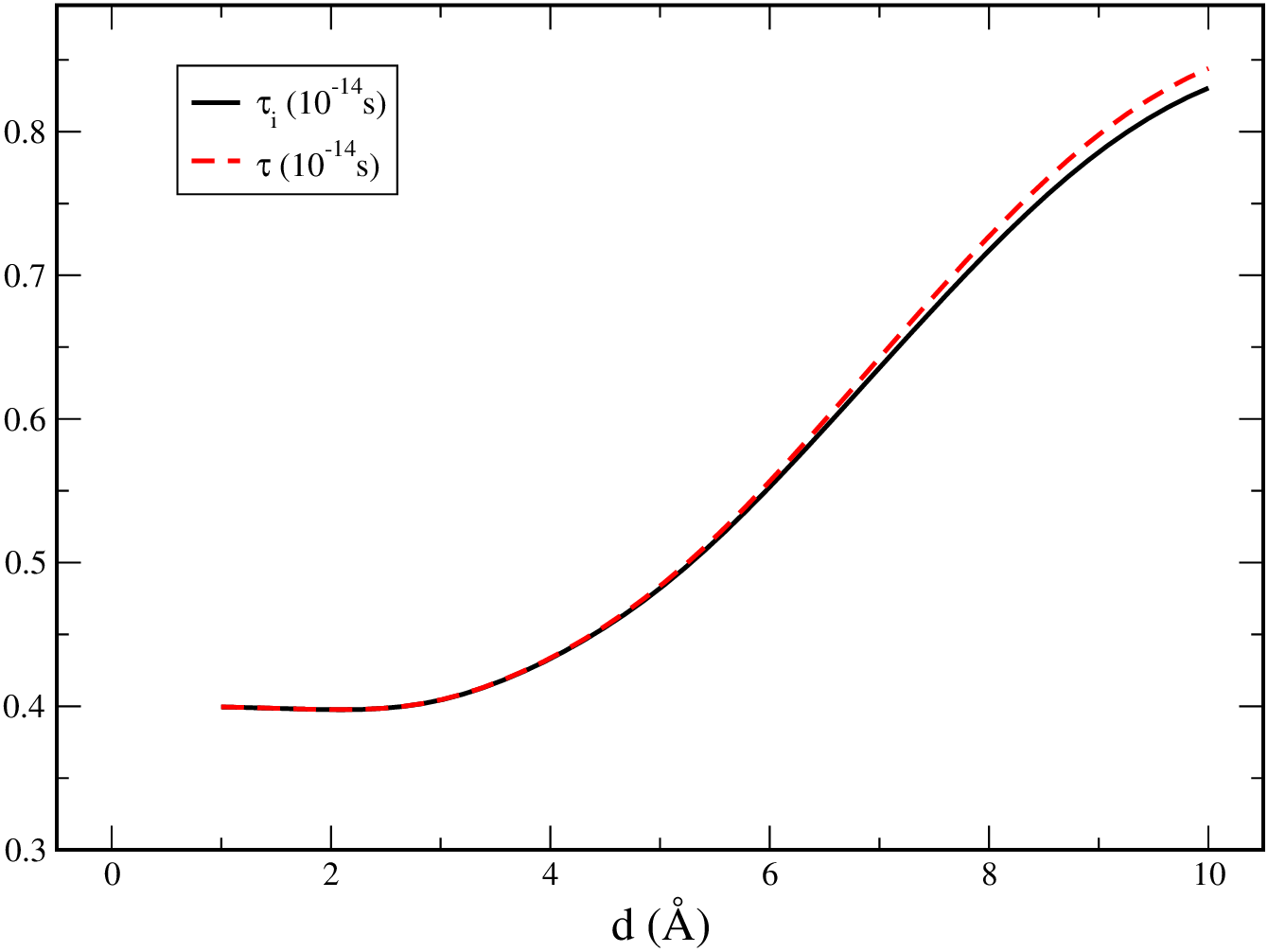}
\caption{The mean arrival time of electrons versus the width of the barrier, without the spin-dependent term (solid curve) and with the spin-dependent term (dashed curve), for a symmetric Gaussian packet passing through a barrier in the x-direction.}
\label{fig: mean-width-b}
\vspace*{.45cm}
\end{figure}

Fig. \ref{fig: mean-width-b} shows the mean arrival time of electrons at the given location, versus width of the barrier. From this figure, it follows that the spin-dependent term increases the mean arrival time ($\tau > \tau_i$) and the quantity $\tau-\tau_i$ is increased with the width of the barrier. 

\begin{figure}
\centering
\includegraphics[width=10cm,angle=0]{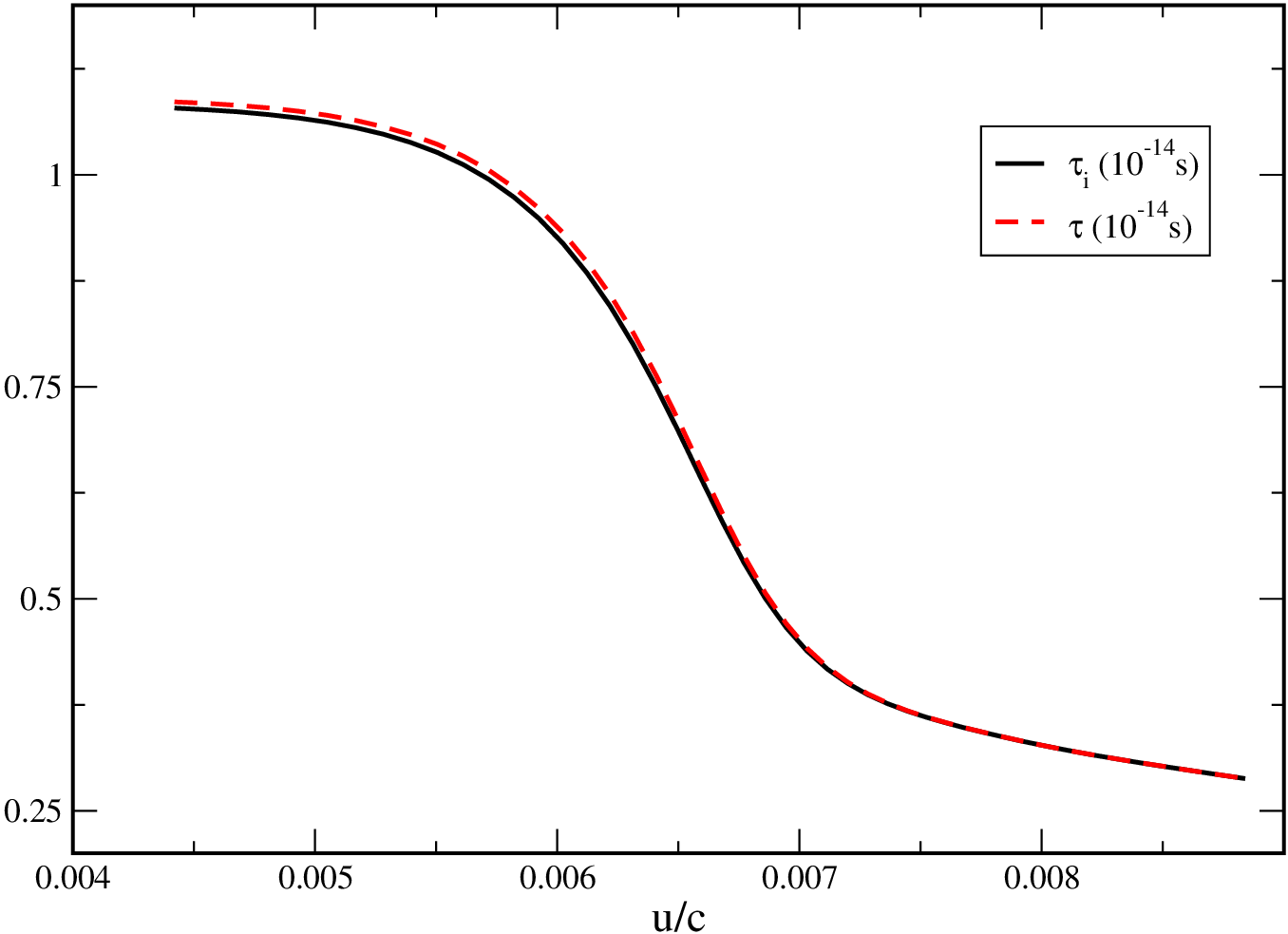}
\caption{The mean arrival time of electrons versus the ratio of the group velocity to the light velocity, without the spin-dependent term (solid curve) and with the spin-dependent term (dashed curve), for a symmetric Gaussian packet passing through a barrier in the x-direction.}
\label{fig: mean-speed-b}
\end{figure}

Fig. \ref{fig: mean-speed-b} shows the mean arrival time of electrons versus the ratio of the group velocity to the light velocity $u/c$ for $d=10$\AA{}. It follows from this figure that the spin-dependent term increases the arrival time ($\tau > \tau_i$) and has very small effect on the mean arrival time at the specified location, and that the quantity $\tau-\tau_i$ is decreased with the initial group velocity. Note that we could not decrease the initial energy (or the initial group velocity) as we want,as in the Fig. \ref{fig: mean-speed-u}, because if $E_0=\frac{1}{2}mu^2 \ll V_0$, then the transmission probability would be very small and no particle can cross the barrier to reach the detector position. 

\begin{figure}
\centering
\includegraphics[width=10cm,angle=0]{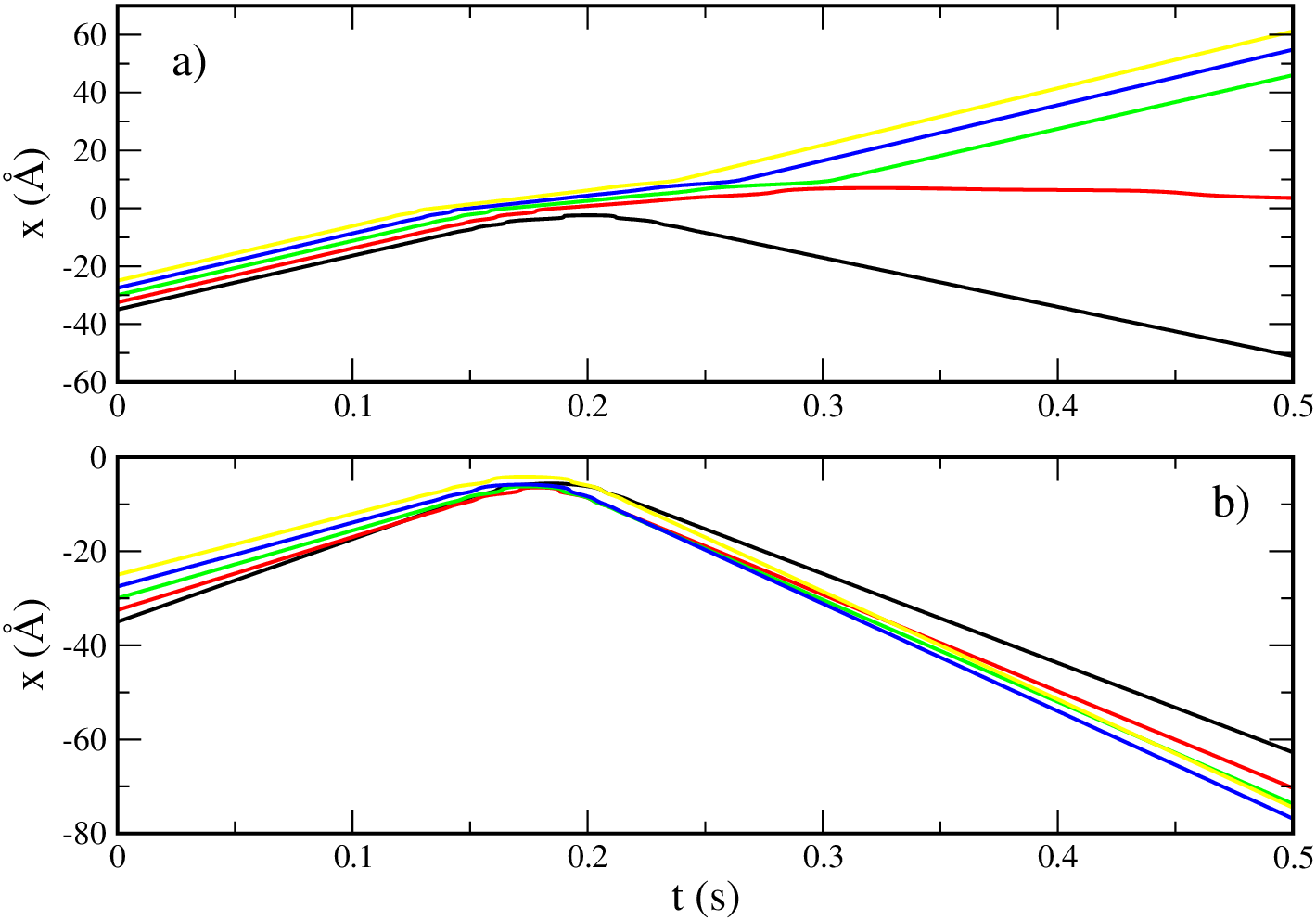}
\caption{Some Bohmian trajectories of electrons in x-t plane with the initial position $(x^{(0)}, y^{(0)}=\sqrt{50 \sigma_0^2-x^{(0)^2}},  z^{(0)}=0)$ a) in the absence of the spin-dependent term and b) in the presence of the spin-dependent term, for a symmetric Gaussian packet passing through a barrier in the x-direction.}
\label{fig: tra_b}
\end{figure}

Fig. \ref{fig: tra_b} shows some Bohmian paths for electrons. One can see that trajectories cross each other (Fig. \ref{fig: tra_b}), when one considers the spin-dependent term. As the Fig. \ref{fig: tra_u} they cross each other for different values of the $y$ coordinate. The spin-dependent term can also change the fate of the individual trajectories. There are some trajectories that don't reach the barrier, with the spin-dependent term, but they cross the barrier without this term. However, ensemble averaging yields the same transmission and reflection probabilities. The spin-dependent term plays a role only in time related measurements.


\section{Summary and discussion}
Spin has a very small effect on the mean arrival time at a given location in the nonrelativistic domain, but it has a significant effect on the fate of individual Bohmian paths. Bohmian paths can cross each other, when one considers the spin-dependent term. The quantity $\tau-\tau_i$ changes with the mass of the arriving particle, with the group velocity of the wave packet and also with the width of the barrier, but, the difference is very small.

\section{Acknowledgment}

We are very indebted to H. Nikolic for helpful discussions and corrections and valuable suggestions. S. V. M. is grateful to J. G. Muga and A. del Campo for providnig useful information.




\end{document}